\documentclass[aps,nofootinbib,twocolumn]{revtex4}
\usepackage{graphicx,amsmath,hyperref,amssymb,bbm}
\def\be{\begin{eqnarray}}
\def\ee{\end{eqnarray}}
\def\no{\nonumber}
\def\vol{{\mathtt{v}}}
\def\Vol{{\mathtt{V}}}
\def\c{{c}}

\begin{document}
\title{ \Large Spinfoam fermions}
    \author{Eugenio Bianchi}
   \affiliation{Centre de Physique Th\'eorique de Luminy\footnote{Unit\'e mixte de recherche du CNRS et des Universit\'es de Provence, de la M\'editerran\'ee et du Sud; affili\'e \`a la FRUMAN.},      Case 907, F-13288 Marseille, EU     }
    \author{Muxin Han}
   \affiliation{Centre de Physique Th\'eorique de Luminy\footnote{Unit\'e mixte de recherche du CNRS et des Universit\'es de Provence, de la M\'editerran\'ee et du Sud; affili\'e \`a la FRUMAN.},      Case 907, F-13288 Marseille, EU     }
       \author{Elena Magliaro}
   \affiliation{Institute for Gravitation and the Cosmos, The Pennsylvania State University, 104 Davey Lab, University Park, PA 16802, USA}
    \author{Claudio Perini}
   \affiliation{Institute for Gravitation and the Cosmos, The Pennsylvania State University, 104 Davey Lab, University Park, PA 16802, USA}
    \author{Carlo Rovelli}
   \affiliation{Centre de Physique Th\'eorique de Luminy\footnote{Unit\'e mixte de recherche du CNRS et des Universit\'es de Provence, de la M\'editerran\'ee et du Sud; affili\'e \`a la FRUMAN.},      Case 907, F-13288 Marseille, EU     }
    \author{Wolfgang Wieland}
   \affiliation{Centre de Physique Th\'eorique de Luminy\footnote{Unit\'e mixte de recherche du CNRS et des Universit\'es de Provence, de la M\'editerran\'ee et du Sud; affili\'e \`a la FRUMAN.},      Case 907, F-13288 Marseille, EU     }
\date{\today}
\begin{abstract}\noindent
We give the definition of a minimal coupling of fermions and Yang Mills fields to the loop quantum gravity covariant dynamics. The coupling takes a surprisingly simple form.   Here we only define the dynamics;  physical implications are considered in a subsequent paper. 
\noindent \end{abstract}
\maketitle
\section{Introduction}

A longstanding difficulty in spinfoam loop gravity has been the problem of coupling fermions to the theory in four dimensions.  This difficulty has been considered a major roadblock for spinfoam theory, and has been repeatedly indicated as a major problem to address. See \cite{Rovelli:2010vv,Rovelli:2010wq} and full references therein.   In the covariant formalism, the difficulty comes from the fact that in four dimensions the fermion action cannot be written in terms of the Plebanski two-form, which is the object naturally defined in the spinfoam formalism.  This complicates matter with respect to the situation in three dimensions, or in the canonical framework, where the triad is directly available as an elementary variable in the theory.  For this reason, while fermion coupling in 3d and in the canonical theory has been developed early, the problem of coupling fermions to the 4d spinfoam theory has so far remained open.  Here we find a way to define a minimal coupling of the spinfoam theory to a chiral fermion field.   We obtain this by observing that the discretisation of the Dirac action can be  manipulated to a form which only contains quantities that have an analog in 4d spinfoams.   

We find that the dynamical coupling of a fermion to quantum gravity is surprisingly simple in the spinfoam formalism, once the suitable  discretisation is found.  This result resonate with similar ones in early canonical loop theory   \cite{MoralesTecotl:1994ns,MoralesTecotl:1995jh}: the naive fermion hamiltonian was found to be just the extension of the simple ``shift" form \cite{Rovelli:1989za} of the gravitational hamiltonian, to the open ends of the gravitational Faraday lines, which represent fermions.  Here we find that in the spinfoams context the fermion dynamics is a natural extension of the gravitational amplitude to spin-networks with spin-$\frac12$ open-ends at the nodes.

We also consider the coupling of a Yang Mills field, a problem that was already studied in \cite{Speziale:2007tg} in 3d and in \cite{Oriti:2002bn} in 4d. In the presence of fermions, it is then easy to add Yang-Mills fields by promoting a global symmetry to a local symmetry. The dynamics can then be generated \`a la Zeld'hovic \cite{Zeldovich:1967fk,Adler:1982uq} from fermion loops. 

The physics that this theory yields is analyzed in a companion paper \cite{Han:2011as}, where we study correlation functions; show that there is a spinfoam analog of PCT symmetry for the fermion fields on spinfoam model, and prove a PCT theoremfor spinfoam fermion correlation functions;  compute the fermion correlation functions and show that they can be given by Feynman diagrams on the spinfoams, where the Feynman propagators can be represented by a discretized path integral of a world-line action along the edges of the underlying 2-complex.

For earlier studies on how to couple fermions to a discretized gravitational field, see \cite{Bander:1987uq, Bander:1988kx,Ren:1988vn,Kawamoto:1991fk}. For detailed studies of fermions in the loop quantum gravity canonical formalism, see  
\cite{Krasnov95,Montesinos:1998dw,Thiemann:1998ys,ThiemannBook}. On fermions in 3d, spinfoams and quantum cosmology and the effect of fermions on Holst gravity, see also \cite{Fairbairn:2006dn,Alexandrov:2008iy, Bojowald:2008nz,Perez:2005pm,Mercuri:2006um}.

\section{Fermions on a 2-complex}

We start in 4d euclidean space for simplicity, and will rotate to the lorentzian theory later on. 
The Dirac action reads 
\be
S_D=i\int d^4x\ \overline\psi_D\! \not{\!\partial} \psi_D
\ee
plus complex conjugate, always understood here and in what follows.
Here we are interested in a chiral spinor $\psi\in\mathbbm{C}^2$ with two complex components $\psi^a, a=0,1$. Its action is the projection of $S_D$ on one of its two helicity components. This reads
\be
S=i\int d^4x\ \overline\psi \sigma^I\partial_I \psi
\label{action}
\ee
where $\sigma^I=(\mathbbm 1, \vec \sigma)$, $\vec\sigma$ are the Pauli matrices and $\overline\psi$ denotes the hermitian conjugate of $\psi$. 

Fix a coordinate system $x=(x^I)$ and discretize spacetime by chopping it into the union of 4-cells $v$. Consider the dual complex, with vertices $v$ (with coordinates $x_v$) connected by (oriented) edges $e$. Call $|v|$ the number of edges bounded by $v$.  Approximate the field $\psi(x)$ by its values $\psi_v=\psi(x_v)$ at  each vertex.  Discretize the derivative on each edge as 
\be
    \psi_{t_e}-\psi_{s_e} \sim  (x_{t_e}^I-x_{s_e}^I)\partial_I\psi
\ee
where $s_e$ and $t_e$ are the source and the target of the edge $e$. 
This gives
\be
      u_e^I\partial_I\psi\sim\frac{\psi_{t_e}-\psi_{s_e}}{l_e}
\ee
where $l_e=|x_{t_e}^I-x_{s_e}^I|$ is the length of the edge $e$ and $u_e^I=(x_{t_e}^I-x_{s_e}^I)/l_e$ is the unit vector parallell to $e$.  The action \eqref{action} can then be discretized as a sum over 4-cells: $S\to\sum_v S_v$. 
\be
S_v
\sim\frac{4i\Vol_v}{|v|}\sum_{e\in v} \overline\psi_e u_{e}^I\sigma_I \frac{\psi_{v+e}-\psi_v}{l_e},
\ee
where the sum is over the edges bounded by $v$, $\Vol_v$ is the volume of the 4-cell $v$, and the factor $\frac4{|v|}$ is included to take into account the fact that multiple edges over-count the derivative. The second term above cancels in subtracting the complex conjugate. Now consider the 3-cell $\tau_e$ dual to the edge $e$. Assume this is orthogonal to the edge (as in a Voronoi cellular complex). Each 4-cell $v$ can be partitioned into the union of $|v|$ pyramids with base $\tau_e$ and hight $h_e=l_e/2$. The 4-volume of these is $\frac14 h_e {\vol}_e$ where ${\vol}_e$ is the 3-volume of $\tau_e$. Using this, 
\be
S_v\sim\frac{ i }{2} \sum_{e\in v} \ \overline\psi_e \vol_e \sigma_e\psi_{v+e},
\ee
where
\be
\sigma_e \equiv \sigma^Iu_{eI}
\ee
is the $\sigma$-matrix ``in the direction of the edge $e$". This can be written as 
\be
\vol_e\sigma_e \equiv \sigma^I  \int_{\tau_e} \epsilon_{IJKL}\  e^J\wedge e^K\wedge e^L, 
\label{sigma}
\ee
where $e^I=dx^I$ is the tetrad one-form.  Notice now that each term of the sum depends only on edge quantities. This suggest to consider writing the full discretized action as a sum of edge terms
\be
S= i\sum_{e}\    \overline\psi_{s_e}\vol_e \sigma_e  \psi_{t_e}.
\label{da}
\ee
More in general, say that the collection of edges $e$, together with the data $(\vol_e, u_{eI})$ approximate a flat metric at a scale $a$ if  \cite{Ashtekar:1992tm}
\be
     \int d^4x\  \delta^\mu_I\omega_\mu^I(x)
 = \sum_e \vol_e  \int_e\omega^Iu_{eI}
 \ee
for any quadruplet of one-forms $\omega^I=\omega_\mu^Idx^\mu$ that varies slowly at the scale $a$.  Then \eqref{da} is a discretization of the Weyl action if the two-complex approximates a flat metric. Equation \eqref{da} is a very simple expression that discretizes the fermion action.%
\footnote{Fermions on a lattice suffer from the fermion-doubling problem and the related chiral anomaly.  Because of the absence of a regular triangulation and the integration on the gravitational variables that we introduce below, however, here fermions are essentially on a random lattice, where the (obvious) species doubling problem does not arise.}  
Its simplicity recalls the simplicity of the free particle hamiltonian on a graph \cite{Rovelli:2009ks}. It could have also been directly guessed from \eqref{sigma} and the form
\be
S=i\int  \overline\psi\, \sigma^I d \psi\wedge e^J\wedge e^K\wedge e^L\ \epsilon_{IJKL}
\label{action2}
\ee
of the action \eqref{action}.

In view of the coupling with gravity, and in order to better understand boundary states, it is convenient to move from vertex variables to edge variables. Let $x_e$ be the intersection between $e$ and $\tau_e$ and introduce edge variables $\psi_{e}=\psi(x_e)$. In the approximation in which we are working, where the second derivative of the field can be neglected, $ \psi_{e}\sim\frac12(\psi_{s_e}+\psi_{t_e})$. 
Using this and the fact that the sum of quadratic terms averages to zero, we can write 
\be
S = \frac{i}4\sum_{ve}\  \overline{\psi}_{v}  \vol_e\sigma_e \psi_{e}.
\label{e} 
\ee 
where the sum is over all couples of adjacent vertices-edges.
Then observe that (always in this approximation) we can express the vertex fermion as an average over the corresponding boundary edge fermions: $\psi_v=\frac{1}{|v|}\sum_{e'\in v}\psi_{e'}$. This gives 
\be
S\sim \sum_{v} \frac{i}{4|v|}\sum_{e,e'\in v}\, \overline{\psi}_{e'} \vol_e \sigma_e\psi_{e},
\ee 
Thus, we consider the discretization of the fermion action defined by associating to each 4-cell $v$ with boundary fields $\psi_e$ the action
\be
S_v= i\sum_{ee'}  \, \overline{\psi}_{e'} \vol_e\sigma_e \psi_{e}.
\label{questa}
\ee
where we have assumed here for simplicity that all vertices have the same valence and we have absorbed a constant in a redefinition of the field. This is an expression that can be used to couple the fermion to quantum gravity.

\section{Quantum fermion field on a 2-complex}

Consider the fermion partition function of a 2-complex characterized by the quantities $(\vol_e,\sigma_e)$.  This will be given by the Berezin integral
\be
Z=\int D\psi_e\ e^{iS}.
\label{bere}
\ee
Choose the integration measure to be 
\be
 D\psi_e = \frac{1}{\vol^2_e}\ d\psi_ed\overline\psi_e \  e^{-\vol_e\overline\psi_e\psi_e},
 \label{measure}
\ee
which realizes the scalar product at each edge, seen as a boundary between two 4-cells, and where we interpret the field $\psi_e$ as an anticommuting variable, in order to take Pauli principle into account. The volume $\vol_e$ in the exponent is needed for dimensional reasons and to keep into account the fact that in the classical theory 
\be
\langle \psi |\psi'\rangle = \int d^3x \sqrt{q}\ \overline{\psi(x)} \psi'(x)
\ee
contains the 3-volume factor $\sqrt{q}$ \cite{Rovelli:2009ks}. 
The definition of Berezin integral is that the only non vanishing integral is 
\be\label{bre}
\int d\psi d\overline\psi \ \psi^a\overline\psi{}^b \psi^c\overline\psi{}^d
= \epsilon^{ac}\epsilon^{bd}
= (\delta^{ab}\delta^{cd}-\delta^{ad}\delta^{bc}).
\ee
Expand the vertex action in Taylor series
\be
e^{iS_v}&=&1+\sum_{ee'}  \overline{\psi}_e \vol_e\sigma_e \psi_{e}\\
\nonumber 
&& +
\sum_{e_1e_2e_3e_5} (\overline{\psi}_{e_1} \vol_{e_2}\sigma_{e_2} \psi_{e_2})(\overline{\psi}_{e_3}  \vol_{e_4}\sigma_{e_4} \psi_{e_4})+...
\ee
The series stops because there can be at most 4 fermions (two $\psi$ and two $\overline\psi$) per edge.  Each edge integration in \eqref{bere} gives zero unless on the edge there is either no fermion, or a term $\overline\psi_e\psi_e$, or a term $\overline\psi_e\psi_e\overline\psi_e\psi_e$. The volume factors in the measure cancel those in the vertex action. The result is that \eqref{bere} becomes a sum of terms, each being the product of traces of the form 
\be
       Z= \sum_{\{c\}} \prod_\c A_{\c}
       \label{Z}
\ee
where $\{\c\}$ is a collection of oriented cycles $\c$, each formed by a closed sequence of oriented edges on the 2-complex: $\c=(e_1, ..., e_N)$, and each trace is 
\be
A_{\c}=(-1)^{|\c|} \ Tr[\sigma_{e_1}...\sigma_{e_N}],
\label{trace}
\ee
where $|\c|$ is the number of negative signs from \eqref{bre}.
For every edge, there cannot be more than two fermionic lines (Pauli principle) and if there are two lines, these are anti-symmetrized (by \eqref{bre}).  Explicitly, each term reads
\be
      A_{\c}={(-1)}^{\scriptscriptstyle|\c|}\ Tr[\sigma_{I_1}...\sigma_{I_n}]\  u_{e_1}^{I_1}... u_{e_n}^{I_n}\label{Agamma}
\ee
and is therefore a Lorentz-invariant contraction of the normals to the 3-cells.%
\footnote{If we do not include the volume factors in  \eqref{measure}, then \eqref{trace} is replaced by 
\be
A_{\c}=(-1)^{|\c|}\ \vol_{e_1}... \vol_{e_N} \ Tr[\sigma_{e_1}...\sigma_{e_N}].
\label{trace2}
\ee This alternative will be studied elswhere.}

\section{Fermions in interaction with gravity}

Let us now return to the Lorentzian framework, and come to our first key technical observation. Assume that all edges are time-like.\footnote{Spacelike edges and timelike 3-cells \cite{Perez:2000ep,Conrady:2010fk} will be considered elsewhere.}  Consider an $SL(2,\mathbbm C)$ matrix $g_e$ that rotates the unit vector $u^I=(1,0,0,0)$ into the vector $u_e^I$ (the phase is irrelevant, and can be fixed by requiring that $g_e$ is a pure boost)
\be
u_J\Lambda^J_{g_e}{}_I= u_{eI},
\ee
where $\Lambda_g$ is the vector representation of $SL(2,\mathbbm{C})$. Recall the transformation properties of the $\sigma^I$ matrices, 
\be
\Lambda_g^I{}_J\sigma^J=g^\dagger \sigma^I g,
\ee
and observe that 
\be
           u_e^I\sigma_I=g_e^\dagger  \sigma^0 g_e =g_e^\dagger g_e, 
\ee 
because $\sigma^0=1$.  Therefore we can write the discretized action \eqref{da} in the form
\be
S=i\sum_{e} \vol_e\,  \overline{\psi}_{s_e} g_e^\dagger g_e\psi_{t_e}.
\label{afff}
\ee
and the vertex action as 
\be
S_v= i\sum_{ee'} \vol_e  \, \overline{\psi}_{e'} g_e^\dagger g_e \psi_{e}.
\label{questaqui}
\ee
What is the geometrical interpretation of these $SL(2,\mathbbm C)$ group elements? 

Consider a single chiral fermion. This can be thought as a quantum excitation of a single mode of a fermion field, or as a fermionic particle with spin $\frac12$ at one space point.\footnote{On the notion of particle in the absence of Poincar\'e invariance, see \cite{Colosi:2004vw}.}  The fermion $\psi$ transforms in the fundamental representation $H^\frac12\sim\mathbbm {C}^2$ of $SU(2)$. The vector space $\mathbbm {C}^2$ is also the carrier space $H^{(\frac{1}{2},0)}$ of the fundamental representation of  $SL(2,\mathbbm {C})$, and this determines the Lorentz transformation properties of the fermion. But $H^\frac12$ is a unitary representation, while $H^{(\frac{1}{2},0)}$ is not; in other words, $\mathbbm {C}^2$ can be equipped with a scalar product which is $SU(2)$ invariant, but this scalar product is not $SL(2,\mathbbm {C})$ invariant.%
\footnote{If we fix a basis in $\mathbbm {C}^2$, the scalar product is given by $\langle \psi|\phi\rangle=\overline \psi{}^a \phi^b \delta_{ab}$ and the tensor $\delta_{ab}$ is invariant under $SU(2)$ but not under $SL(2,\mathbbm {C})$.}  This becomes particularly transparent if we write $\langle \psi|\phi\rangle=\overline \psi \sigma^0 \phi$, which shows that the scalar product is the time component of a 4-vector $\overline \psi \sigma^I \phi$. The Dirac action exploits this  dependence by constructing a Lorentz scalar contracting the Lorentz vector $s^I=\overline \psi \sigma^I d\psi$ with the covector $u_I$ giving the direction of the derivative, where $\partial_I=d(u_I)$.  

At the light of this discussion, the geometrical interpretation of the matrices $g_e$ is clear. In the discretized theory, a Lorentz frame is determined at each 3-cell  by the 4-normal to the 3-cell, which is the direction along which the derivative is computed.  The matrices $g_e$  parallel-transport the fermion from the fixed reference frame at the center of the 4-cell to a frame at the center of a 3-cell, namely at the boundary of the 4-cell, where the normal to the 3-cell is oriented in the time-direction $(1,0,0,0)$. The scalar product that defines the action is taken in that frame. The construction is Lorentz invariant, since the preferred frame of the scalar product is determined by the normals along which the variation of the fermion is computed. 

Now, \eqref{questaqui}
 is of particular interest for generalizing the discretization to a curved spacetime. In fact, we can discretize a curved geometry in terms of flat 4-cells glued along flat 3-cells, as in Regge calculus. Curvature is then confined on 2-cells. This implies that the holonomy of the spin connection around a 2-cell can deviate from unity. Therefore in general on a curved space there is no way of choosing a reference frame in each 4-cell which is parallel transported to itself across all 3-cells. In other words, the parallel transport from $s_e$ to $e$ may be different from the parallel transport from $t_e$ to $e$. Therefore the generalization of the fermion action to a curved spacetime, discretized \`a la Regge, can be obtained replacing \eqref{afff} with
\be
S=i\sum_{e} \vol_e\,  \overline{\psi}_{s_e} g_{es_e}^\dagger g_{et_e}\psi_{t_e}.
\label{afff}
\ee
where $g_{ev}$ is the holonomy of the spin connection from a coordinate patch covering the (flat) 4-cell $v$ to one covering the (flat) 3-cell $e$. Equivalently, replacing the vertex aplitude \eqref{questaqui}
by 
\be
S_v=i\sum_{e'e} \ \vol_e\ \overline{\psi}_{e'}\  g_{es_e}^\dagger g_{et_e}\psi_{e}.
\label{esta}
\ee
The fermion partition function on the 2-complex representing a curved spacetime, is therefore as before, with the only difference that the amplitude \eqref{Agamma} of each cycle is  replaced by 
\be
      A_{\c}=(-1)^{|\c|}\ Tr[g_{e_1v_1}^\dagger g_{e_1v_2}...g_{e_nv_n}^\dagger g_{e_nv_1}].\label{Agamma22}
\ee
where $(v_1,e_1,v_2,e_2,..., v_n,e_n)$ is the sequence of vertices and oriented edges crossed by the cycle $\c$. This can be written in a form more easy to read by defining
\be
    g^*=(g^{-1})^\dagger=-\epsilon g^\dagger \epsilon.
\ee
Notice that for a rotation $g^*=g$ while for a boost $g^*=g^{-1}$. Using this, 
\be
      A_{\c}=(-1)^{|\c|}\ Tr[g_{v_1e_1}^* g_{e_1v_2}...  g_{v_ne_n}^* g_{e_nv_1}],\label{Agamma2}
\ee
where the sequence of vertices and edges is in the cyclic order. 
The full partition function \eqref{Z} becomes
\be
Z=\sum_{\{\c\}} 
\prod_c (\raisebox{1pt}{-}1)^{|\c|}\  \chi^{\frac12}\!\Big(\!\prod_{e\in\c}(g^{*}_{s_ee} g_{et_e})^{\epsilon_{e\c}}\!\Big),
\ee
where $\chi^{\frac12}(g)$ is the character in the fundamental representation of $SL(2,\mathbbm C)$, and $\epsilon_{e\c}=\pm 1$ according to whether the orientations of the edge and the cycle match. 

The action \eqref{afff} and the amplitude \eqref{Agamma2} are of particular value for coupling the fermion field to quantum gravity because they depend on the geometry only via the two quantities $\vol_e$ and $g_{ve}$, which are precisely the quantities that appear in the gravitational spinfoam amplitude. This we do in the next section.

\section{Coupling to quantum gravity}

The spinfoam partition function of pure gravity on a 2-complex can be written in the holonomy representation \cite{Magliaro:2010ih,Bianchi:2010mw} in the form \cite{Rovelli:2010vv}
\begin{eqnarray}
\label{int2}
Z&=& \int_{SL(2,\mathbbm C)}dg_{ve}\int_{SU(2)}dh_{e\!f}\;
 \sum_{{j_{\!{}_f}}}
 \prod_{f} d_{j_{\!{}_f}}
  \\ \nonumber
&&  
\chi^{\gamma j_{\!{}_f}\!,{j_{\!{}_f}}}\!\Big(\!\prod_{e\in\partial f}(g_{es_e} h_{e\!f} g^{-1}_{et_e})^{\epsilon_{l\!f}}\!\Big) \prod_{e\in\partial f}\chi^{j_{\!{}_f}}\!(h_{e\!f}).
\end{eqnarray}
Now, the second key observation of this paper is that the $SL(2,\mathbbm C)$ matrices $g_{ev}$ in this expressions can be identified with the  $SL(2,\mathbbm C)$ matrices $g_{ev}$ in  \eqref{afff} and \eqref{Agamma2}. Indeed, first, they have the same geometrical interpretation as parallel transport operators from the edge to the vertex; and second, more importantly,  the asymptotic analysis of the vertex amplitude in \cite{Barrett:2009mw} shows that the saddle point approximation of the integral is on the value of $g_{ev}$ that rotates the (arbitrary) Lorentz frame of the 4-cell into a Lorentz frame at the 3-cell where the time direction is aligned with the normal of the 3-cell, which is precisely the geometry described in the previous section. Therefore in the limit in which we move away from the Planck scale, these group elements take precisely the value needed to yield the fermion action. 

The obvious ansatz for the dynamics in the presence of fermions, is therefore to replace \eqref{int2} by 
\begin{eqnarray}\label{int3}
Z& = & \sum_{\{\c\}} \sum_{{j_{\!{}_f}}}  \int_{SL(2,\mathbbm C)}  dg_{ve}\int_{SU(2)}  dh_{e\!f}\nonumber \\ 
&&
 \prod_{f} d_{j_{\!{}_f}}\ 
\chi^{\scriptscriptstyle\gamma j_{\!{}_f}\!,{j_{\!{}_f}}}\!\Big(\!\!\prod_{e\in\partial f}\!(g_{es_e} h_{e\!f} g^{-1}_{et_e})^{\epsilon_{l\!f}}\!\Big)
 \ \prod_{e\in\partial f}\chi^{j_{\!{}_f}}\!(h_{e\!f})
 \nonumber \\ 
&&
\prod_c (-1)^{|\c|}\ \chi^{\frac12}\!\Big(\!\prod_{e\in\c_n}(g_{es_e}g^{\dagger}_{et_e})^{\epsilon_{ec}}\!\Big).
\end{eqnarray}
where $\{\c\}$ labels families of worldlines running along the edges of the foam. The sum is over all families that do not overlap more than once. Notice that the definition of the $*$ in this expression depends on the choice of a specific $SU(2)$ subgroup at each edge, but this dependence drops from the total expression, because of the $SL(2,C)$ integrations, precisely as discussed in \cite{Rovelli:2010ed}. Therefore Lorentz invariance is implemented in the bulk.

This expression defines a quantum theory of gravity interacting with fermions. 

In the next section we express this amplitude in a more conventional local form, in the spin network basis, and we write explicitly the fermion vertex amplitude. For this, we need to discuss the fermion states.

\section{Fermion states} 

Since we are dealing with a quantum field theory, we need the (fermionic) Fock space.  By Pauli principle (or assuming canonical anticommutation relations), this includes only antisymmetric states.  Since there are at most two orthogonal states in $H^{\frac12}$, there are at most two particles with all other quantum numbers equal. A two-particle state is necessarily the singlet, namely the spin-zero component of $H^\frac12\otimes H^\frac12$. Thus, the Fock space has the structure 
\be 
F=\mathbbm{C}\oplus H^\frac12 \oplus A(H^\frac12\otimes H^\frac12)= \mathbbm{C}\oplus H^\frac12 \oplus \mathbbm{C}.\ee 
A basis in this state is given by $|c\rangle$, where $c=\emptyset,+,-,2$ indicates the no-fermion state, the one-fermion states with spin up and spin down, and the two-fermion singlet.

States in $F$ can be conveniently represented as holomorphic Berezin functions $f({\psi})$ of an anticommuting Grassmann variable in $H^\frac12$. Taylor expanding the state gives 
\be
f({\psi})=c_\emptyset+c_a{\psi}^a+c_2\, \epsilon_{ab} {\psi}^a{\psi}^b
\ee
where the last term does not vanish because of the anticommutation properties of ${\psi}^a$ and the Taylor series stops at the second term, for the same reason. The norm of the state is 
\be
|f|^2:= |c_\emptyset|^2 + \overline c_a c_a+ |c_2|^2. 
\ee
Then the quadruple $(c_\emptyset,c_a,c_2)\in \mathbbm C^4\simeq F$. The Berzin integral used in the previous section is simply a way to write this Fock space and its scalar product.

Now combine this fermionic Fock space with the kinematics of quantum gravity.  Fermions must reside on the nodes $n$ of a graph, like in lattice gauge theory. Thus we assign a copy of the Fock space $F$ to each node of the graph.  Therefore the states of the gravity+fermion theory live on the space $(\otimes_l L_2[SU(2)])\otimes(\otimes_n F)$, divided by the gauge action of $SU(2)$ at each node. 
We can write states as $\Psi(h_l,{\psi}_n)$, where $l$ labels the links of the graph and $n$ the nodes.

The spin networks that  form a basis of this state are a simple generalization of the pure gravity spin networks. As before, it is convenient to choose an intertwiner basis at each node $n$ that diagonalizes the volume of the node $n$, and label it with the volume eigenvalue $\vol_n$. That is $|j_l, \vol_n\rangle$. In the presence of fermions, spin networks carry an extra quantum number $c_n$ at each node, which labels the basis $|c\rangle$ in the Fock space at the node.  That is: $|j_l, {\vol}_n, c_n\rangle$. At each $v$-valent node $n$ bounded by links with spins $j_1,...,j_v$, the intertwiner ${\vol}_n$ is an invariant tensor in the tensor product of the $v$ representations $j_1,...,j_v$ if $c_n=\emptyset$ or $c_n=2$.  But it is an invariant tensor in the tensor product of the $v+1$ representations $j_1,...,j_v,\frac12$ if $c_n=\pm$. In this case, the intertwiner couples the spinor to the gravitational magnetic indices.

It is convenient to use a Fock notation, as follows. Write $|j_l, {\vol}_n\rangle$ to denote the state where all the nodes are in the Fock vacuum state, namely the state $|j_l, {\vol}_n, c_n=\emptyset\rangle$. A state with all nodes in the vacuum except for a single node $n$ with a one-particle state with spin $a$ is indicated as $|j_l, {\vol}_n,(n,a)\rangle$. A state with $N$ particles in the nodes $n_1,..., n_N$ with spins $\pm_n$ is indicated by  
$|j_l, {\vol}_n, (n_i,\pm)\rangle$ with $i=1,...,N$. Indicate two-particle singlets as $|j_l, {\vol}_n, (n,+),(n,-)\rangle$, namely as two fermions with opposite spins on the same node. This is the boundary kinematical state space of the theory. 

Now, the spinfoam partition function for pure quantum gravity on a 2-complex in the spin network basis can be written in the form
\be
Z=  \sum_{{j_{\!{}_{\!f}}}\!,\vol_e} \prod_f d_{j_{\!{}_{\!f}}}\ \prod_v A_{v}(j_{\!{}_{\!f}},\vol_e)
\label{int12} 
\ee
where we have chosen a basis in each intertwiner space that diagonalizes the volume, and called $\vol_e$ the quantum numbers as well as the eingevalues of this basis.\footnote{Additional quantum numbers are needed in the case of degeneracy.} The vertex amplitude can be written in the form
\be
A_{v}(j_{\!{}_{\!f}},\vol_e) =  \int dg'_{ev}\  A_{v}(j_{\!{}_{\!f}},\vol_e,g_{ev}),
\label{int122} 
\ee
where (denoting by $l$ and $n$ the links and nodes determined by the intersection of a small 3-sphere surrounding $v$ with the faces $f$ and the edges $e$ of the 2-complex $\cal C$)
\be
A_{v}(j_l,\vol_n,g_n)\!=\! \!
\int dk_l\;  \Psi_{\!j_l\!,v_n}\!(k_l)
\prod_l  \chi^{\gamma j_l\!,j_l}(k_l\,g_{s_l}g^{-1}_{t_l}\!).
\label{vaaa}
\ee
Let us now write the gravity+fermion partition function in the local form 
\be
Z= \sum_{{j_{\!{}_{\!f}}\vol_e}}\!\int  D\psi_{e} \prod_f \!d_{j_f}\prod_v \! A_{v}(j_f,\vol_e,\psi_e)
\label{fff} 
\ee
where the vertex amplitude is
\be
A_v(j_l,{\vol}_n,{\psi}_n)
\!=\!\int \! dg'_n A_v(j_l,{\vol}_n,g_n)\ 
e^{iS_v(g_n,{\vol}_n,{\psi}_n)}
\label{vf}
\ee
and the fermion action $S_v$ is defined in \eqref{esta}. If we expand the vertex amplitude in powers of $\psi_e$, we obtain 
\be
A_v(j_l,{\vol}_n,{\psi}_n)
&=&A_v(j_l,{\vol}_n)+
A_v(j_l,{\vol}_n,(n,a)){\psi}^a_n
\\ \nonumber   && +A_v(j_l,{\vol}_n,(n,a),(m,b))
\overline{\psi}{}^a_n {\psi}^b_m+... 
\ee
The first term of this expansion gives the case with no fermions on the boundary of the vertex, and reduces precisely the pure gravity amplitude $A_{v}(j_l,{\vol}_n)$.  If there is a single fermion, $A_{v}(j_l,{\vol}_n,(n,A))=0$,
which expresses fermion number conservation. The same for any odd number of fermions on the boundary.  If there is a $\psi$ and a $\overline\psi$ on two separate nodes, we obtain
\begin{eqnarray}
&&\hspace*{.5cm}A_{v}(j_l,{\vol}_n,(n,a),(m,b))=
\\ \no &&\hspace*{1.5cm}
\vol_n \int_{SL(2,\mathbbm{C})}\hspace{-2mm} dg'_e\ A_v(j_l,{\vol}_n,g_n)\ (g_ng_m^{\dagger})_{ab}.
\label{va2}
\end{eqnarray}
For two $\psi$ and two $\overline\psi$ on the boundary, 
\begin{eqnarray}
 A_{v}(j_l,{\vol}_n,(n_i,a_i))&=&
\int_{SL(2,\mathbbm{C})} \hspace{-6mm} dg'_e\ 
A_v(j_l,{\vol}_n,g_n)\ \vol_{n_2} \vol_{n_4}\ 
\\ \no &&\hspace*{-3cm}
\left((g_{n_1}g_{n_2}^{\dagger})_{a_1a_2}
(g_{n_3}g_{n_4}^{\dagger})_{a_3a_4}-
(g_{n_1}g_{n_4}^{\dagger})_{a_1a_4}
(g_{n_2}g_{n_3}^{\dagger})_{a_3a_2}\right)
\label{vmany}
\end{eqnarray}
and so on.\footnote{But if we integrate away the vertex fermions $\psi_v$ quantum mechanically instead than solving them away classically as we have done above, the Pauli principle would limit the number of lines crossing the vertex to two, and there are no other terms.}  This defines the vertex amplitude.

The Berezin integration has the only effect of connecting the free indices of the cell-propagators in \eqref{va2} and to cancel the volume factors. This gives the fermion contribution to the spinfoam amplitude as a sum over sequences of contractions of $(g_{ev}g_{e'v}^{\dagger})_{ab}$ terms
\be
Z=\hspace{-.6em} \sum_{{j_{\!{}_{\!f}}\vol_e\{\!\c\!\}}}\!\int \! dg'_{ev} \prod_f \!d_{j_f}\!\prod_v \! A_{v}({\scriptstyle j_f,\vol_e,g_{ev}}\!)\prod_{c}A_{\c}({\scriptstyle g_{ev}}),
\label{ff} 
\ee
which is equivalent to \eqref{int3}. 

Notice that the dependence of the Lorentz frame of the boundary drops in gluing the individual group elements. A local Lorentz transformation is a transformation in each vertex $v$, which affects the holonomies as
\be
        g_{ve}\to \Lambda_vg_{ve},
\ee
which leaves the amplitude invariant. 

The net effect of the fermions is simply to add fermion world-lines over the foam. The weight of each world-line is a contraction of spin-connection group elements along the world-line, taken in the fundamental representation.  The worldlines than run over the foam carry a $j=\frac12$ representation, and couple to the intertwiners at the edges. They overlap at most twice and where they overlap, they run in the $j=0$ representation. 

\section{Yang-Mills fields}

Suppose now that the fermion lives in fundamental representation of a compact group $G$. Then the above theory is invariant under global $G$ trasformations.  To make it invariant under local gauge transformations we can introduce a group element $U_{ve}\in G$ associated to each wedge $(v,e)$, and replace \eqref{da} by 
\be
S= i\sum_e \  \overline{\psi}_{s_e} U^{\dagger}_{s_ee} \vol_e \sigma_eU_{t_ee}{\psi}_{t_e}\label{pippopippo}.
\ee 
The quantum kinematics on the boundary is then evident: spinfoams carry representations of $SL(2,\mathbbm{C})$ and intertwiners at the nodes have a possible extra leg representing fermions in (antisymmetric products of) the fundamental representation of $SL(2,\mathbbm{C})\times G$. 

What is the dynamics? One possibility of obtaining it is simply to keep only the gravity and fermion terms in the action.  The Yang-Mills action is then generated by the one-loop radiative corrections to the fermion action in the Yang-Mills field, as suggestd by Zel'dovich \cite{Zeldovich:1967fk, Adler:1982uq}.  

That is, we can take the dynamics to be defined by 
\be
Z= \sum_{{j_{\!{}_{\!f}}\vol_e}}\!\int  d\psi_{e}\int  dU_{e} \prod_f \!d_{j_f}\prod_v \! A_{v}(j_f,\vol_e,\psi_e,U_e)
\label{fff} 
\ee
where the gravity+Yang-Mills+fermion vertex amplitude is
\be
A_v(j_l,{\vol}_n,{\psi}_n,U_{n})
\!=\!\!\!\int \! dg'_n A_v({\scriptstyle h_l,{\vol}_n,g_n})
\ e^{iS(g_n,{\vol}_n,{\psi}_n,U_{v})}
\label{vf}\!\!.
\ee
In the holonomy representation, this gives
\begin{eqnarray}
\label{int4}
Z& = & \sum_{\{\c\}} \sum_{{j_{\!{}_f}}}  \int_{SL(2,\mathbbm C)}  dg_{ve}\int_{SU(2)}  dh_{e\!f}\; \int_{G} dU_{ve} \nonumber \\ 
&&
 \prod_{f} d_{j_{\!{}_f}}\ 
\chi^{\scriptscriptstyle\gamma j_{\!{}_f}\!,{j_{\!{}_f}}}\!\Big(\!\!\prod_{e\in\partial f}\!(g_{es_e} h_{e\!f} g^{-1}_{et_e})^{\epsilon_{l\!f}}\!\Big)
 \ \prod_{e\in\partial f}\chi^{j_{\!{}_f}}\!(h_{e\!f})
 \nonumber \\ 
&&
\prod_c (-1)^{|\c|}\ \chi^{\frac12}\!\Big(\!\prod_{e\in\c_n}(g_{es_e} \! U_{es_e}\! U^{\dagger}_{et_e} g^{\dagger}_{et_e})^{\epsilon_{ec}}\!\Big).
\end{eqnarray}

This  expression defines a minimally-coupled spinfoam formulation of the Einstein-Weyl-Yang-Mills system.   It can be used to compute all quantum gravity amplitudes order by order. The full theory is formally recovered in the limit of large 2-complexes \cite{Rovelli:2010wq, Rovelli:2010qx}.

\section{Concluding remarks}

We have defined a tentative quantum theory of gravity, fermions and Yang Mills fields.  The strategy we have followed in order to couple fermions to quantum gravity is the following.  A spinfoam amplitude can be seen as a definition of the gravitational path integral which keeps into account the quantization of geometry: intermediate states along the evolution are properly a basis in the LQG Hilbert space.   To couple fermions, we have found a discretization of the fermion action which remains valid on a curved spacetime and which is expressed in terms of the same variables that appear in the spinfoam amplitudes. This discretization couples naturally with the spinfoam amplitudes.  As is always the case in quantum gravity, the appropriate way to view this calculation is not as a ``derivation" or a ``quantization", but rather as a heuristic hint,  yielding an ansatz for a definition of the coupled gravity-fermion theory.   The theory appears to have the correct degrees of freedom and the proper symmetries. 

In the Appendix, we sketch an heuristic relation between the amplitude we have found and a single fermionic particle path integral.  We leave several issues open. Among these the possible existence of spacelike edges and timelike 3-cells \cite{Perez:2000ep,Conrady:2010fk}, and the possible presence of the volume factors in the amplitude.

A minimal coupling of fermions to quantum gravity appears to be surprisingly simple.  This naturalness of the fermion dynamics in gravity has been already observed \cite{MoralesTecotl:1994ns,MoralesTecotl:1995jh} and reappears here. A fermion, is essentially an extra ``face" of spin $\frac12$, which is non-local over the 2-complex. At fixed time, it can be seen as a ``non-local" loop that disappears outside spacetime, to reappear far away, like a Wheeler-Smolin ``Kerr-Newman fermion": the picture of fermions as wormholes has been advocated by John Wheeler long ago \cite{Wheeler:1962fk,Sorkin:1977uq}, considered by Lee Smolin \cite{Smolin:1994uz} in the context of loop gravity, and remains intriguing.  On the idea of ``Kerr-Newman" particles, namely the connection between a fermion and a minimal Kerr-Newman \cite{Newman:1965kx,Newman:1965vn} black hole, see \cite{Arcos:2007zr} and references therein. Recall that both fermions and rotating black holes have a gyromagnetic factor $g= 2$, a surprising and intriging coincidence which has been indicated as a hint to a deeper connection \cite{Pfister:2003ys}.

Some of the physical consequence from spinfoam fermion is analyzed in a companion paper \cite{Han:2011as}. However there are many investigations which should be carried out in the future, in order to further understand the physics of matter quantum fields interacting with spinfoam LQG. As an example, the investigation of semiclassical asymptotic expansion of spinfoam model coupling to fermions and Yang-Mills fields is planned in the near future, by generalizing the method developed in \cite{semiclassical,CF,HZ,han}.

\vskip.3cm
\centerline{-----}
\vskip.3cm
Thanks to Thomas Krajewski and Alejandro Perez for numerous useful inputs and exchanges.
\appendix
\section*{Appendix: Mass}

Consider the one-particle sector of the theory defined by the sum over all path that connect an initial one-fermion state to a final one-fermion state on a fixed background geometry, namely at fixed values of $j_f,\vol_e$ and $g_{ev}$.  This is given by the sum of the path on the foam going from the initial to the final point, each weighted by the holonomy along the path.  This yields the fermion propagator on the given geometry.  To provide an intuition about this term, lets sketch the modification of the theory needed to include a mass, even if this does not appear to work well with the spinfoam variables.

A Majorana mass term can be added by replacing \eqref{questa} with
\be
S_v= i\sum_{ee'}\vol_e \overline{\psi}_{e'} \sigma_e(1+im\lambda_{ee'}){\psi}_e \label{pippopippo}
\ee
where $m$ is the mass and $\lambda_{ee'}$ is a geometrical length dependent on the geometry at vertex $v$ and on the two edges $e,e'$. The simplest possibility is to choose 
\be
\lambda_{ee'}= \frac{\Vol_v}{\sqrt{\vol_e\vol_{e'}}}
\ee
which is a measure of the distance between $e$ and $e'$. The difficulty of this term is that the 4-volume is not an easy quantity to write in terms of spinfoam variables.  But for now let us disregard this and get an intuition about the effect of this term on a given geometry. The term gives an additional contribution $\prod_v (1+im\lambda_{ee'})$ to each fermion line. If $m\lambda_{ee'}$ is small, we can write 
\be
\prod_v (1+im\lambda_{ee'})\sim e^{im\sum_v\lambda_{ee'}}=e^{im L}
\ee
where $L$ is the proper length of the fermion line.  This gives the fermion propagator as a sum of paths weighted by the exponential of its mass times the proper length of its world-line. 

\vfill


\end{document}